\documentstyle[prb,aps,epsf]{revtex}
\begin{document}
\draft
\twocolumn[\hsize\textwidth\columnwidth\hsize\csname @twocolumnfalse\endcsname

\title{Transverse Phase Locking for Vortex Motion in
Square and Triangular Pinning Arrays}  
\author{C.~Reichhardt$^1$ and C.J.~Olson$^2$} 
\address{$^1$
Center for Nonlinear Studies,
 Los Alamos National Laboratory, Los Alamos, New Mexico 87545 \\
$^2$ Theoretical and Applied Physics, Los Alamos National Laboratory,
Los Alamos, New Mexico 87545} 

\date{\today}
\maketitle
\begin{abstract}
We analyze   
transverse phase locking for vortex motion in a superconductor 
with a longitudinal DC drive and a transverse 
AC drive. 
For both square and triangular arrays we 
observe a variety of 
fractional phase locking steps in the velocity versus DC drive
which correspond to stable vortex orbits. 
The locking steps are more pronounced for the triangular 
arrays which is due to the fact that the vortex motion has a 
periodic transverse velocity component even for zero transverse AC drive.   
All the steps increase monotonically in width with AC amplitude.
We confirm that the width of some
fractional steps in the square arrays scales as the square of
the AC driving amplitude. 
In addition we demonstrate scaling in the 
velocity versus applied DC driving curves at depinning and on the main step, 
similar to that seen for phase locking in charge-density wave systems.
The phase locking steps are most prominent for commensurate 
vortex fillings where the interstitial vortices form symmetrical 
ground states.
For increasing temperature, the fractional steps are washed out very
quickly, while the main step gains a linear component and disappears at
melting. 
For triangular pinning arrays
we again observe transverse phase locking, with  
the main and several
of the fractional step widths scaling linearly with 
AC amplitude.     
\end{abstract}

\vskip2pc]
\narrowtext

\section{Introduction}
The phenomena of phase locking can occur in systems 
interacting with a periodic potential when
an external 
AC drive is superimposed over a DC drive. 
A resonance or 
frequency matching can occur between the AC frequency and
the internal frequency generated by motion over the
periodic substrate.   
The best known example is 
that of Shapiro steps in AC/DC-driven single  
Josephson-junctions \cite{Shaprio1,Barone2} and 
arrays of Josephson-junctions \cite{ShaprioN3},  
where steps are observed in
the current-voltage characteristics. 
Shapiro like phase locking has also been extensively studied in 
charge-density-wave systems \cite{Thorne4}, spin-density waves \cite{Gruner5} 
and vortices in superconductors with 
periodic substrates \cite{Daldini6,Jung7,VanLook8,Shaprio29}.
In addition, in elastic media moving over
{\it random} disorder, a washboard signal can be generated by the 
periodicity of the elastic media itself 
\cite{Maeda10,Olson11}, allowing Shapiro 
like steps to be observed when an AC drive is applied,
such as in vortex systems with random disorder
\cite{Harris12,Kolton13}. 
In all these systems the AC drive is 
superimposed in the {\it same} direction as the DC drive. 
Another characteristic of Shapiro like 
phase locking steps is that the step width
oscillates with AC amplitude and frequency
between zero and a finite value as 
an $n$th order Bessel function $J_{n}(A)$, where $n = 0$ corresponds to 
the depinning or critical current and $n >0$ corresponds to 
higher order steps.  

Recently a different kind of phase locking was proposed for 
a 2D system of vortices moving through a periodic substrate. 
Here the AC drive is applied in the {\it perpendicular} direction,
transverse to the longitudinal applied DC drive \cite{Phase14}.   
Phase locking steps, distinct from Shapiro steps,
were observed in the velocity versus applied DC drive.
The most pronounced step occurs for DC drive values that allow
a sinusoidal orbit to fit into each square pinning plaquette.
For larger AC amplitudes a number of additional fractional 
steps were also observed.     
In addition, unlike Shapiro steps, 
the width of the depinning and higher order steps 
were shown analytically and in simulation to
increase monotonically with AC amplitude
as $(A/\omega)^2$,
where $A$ is the AC amplitude and
$\omega$ is the AC frequency. 
A similar transverse phase locking
phenomena can also occur for an ordered vortex lattice moving over
random disorder, as was recently demonstrated 
\cite{Kolton215}. In the random disorder case, the steps
differ from those observed for square periodic pinning.
The step width for random disorder is more like a Shapiro step, 
going as $\propto |J_{1}(A)|$.  In addition, for low AC amplitudes, the 
step width grows linearly, rather than quadratically 
as in the case of square pinning.  

There are several open questions for the transverse phase locking in the
square pinning arrays.
It is not known how the additional fractional 
steps scale with AC amplitude, what the vortex orbits look like
along these steps, or what the noise signatures are.
It also not known if there is any scaling at the transition to the
phase locked state  similar to that found in 
charge density wave systems, where critical scaling is known to occur
at depinning and along the transitions into and out of the phase locked
regions \cite{Fisher16,Middleton17,Higgins18}.
The effect of disorder on the transverse phase locking has also not been
investigated. 
The vortex filling 
fraction or applied field in experiments will likely determine 
the step width or whether the steps occur at all.
For certain filling fractions the overall vortex lattice can be disordered
or frustrated \cite{Reichhardt19}, which may destroy the steps. 
In addition typical transport experiments are often
done near $T_{c}$ so the effects of finite temperature on the
fractional and main steps should also be investigated to determine the
feasibility of observing these states experimentally.  

It would also be interesting 
to examine the effects of a transverse 
AC drive on systems in which there is an 
internal transverse oscillation even in the absence of an AC drive. An 
example of such a system is vortex motion through a triangular pinning array.  
Interstitial vortices driven through such an array by a longitudinal
DC drive oscillate back and forth in the transverse direction as they
move in order to pass through each minima of the interstitial potential.
In this case the additional 
internal transverse oscillation may lock with the 
AC transverse force, leading to an enhancement of the transverse phase locking 
similar to that observed for the square case. 
Alternatively, the phase locking may be 
more Shapiro like, such as the steps found in the transverse 
phase locking with random pinning \cite{Kolton215}. 

We focus on the transverse phase locking for vortex
motion in square and triangular pinning arrays
in thin film superconductors. 
Vortex pinning and dynamics 
in thin samples with  periodic pinning have attracted
considerable attention, since the effects of pinning can be 
systematically controlled. These pinning arrays can be created using
an array of 
holes \cite{Hebard20,Metlushko21,Field22,Moschalkov23,Harada24} 
or magnetic dots \cite{Schuller25,Hoffman26}. 
Various 
pinning geometries have also been
created such as square \cite{Metlushko21,Field22,Harada24}, 
triangular \cite{Schuller25}, 
rectangular \cite{Hoffman26}, and Kagome \cite{Laguna27}. 
Pronounced commensurability effects appear as peaks in the critical
current at integer matching fields and fractional fields where the
vortices can form a symmetrical configuration with the pinning array
so that the vortex-vortex interaction energy is reduced. 
These matching configurations have been imaged directly in experiment
\cite{Field22,Harada24}
and also observed in simulations \cite{Reichhardt19}. 
For large pinning sites above the first matching field, 
where there are more vortices than pinning sites, 
multiple vortices can be captured by each pin
up to a saturation point.  Beyond this saturation number, additional 
vortices will be located between the pinning sites in the
interstitial regions. For small pinning sites, the saturation
number is one, so that for all fields above the first
matching field the vortices 
are located at the interstitial regions. Direct imaging and simulations
have shown that these interstitial vortices can form ordered 
commensurate
arrangements \cite{Reichhardt19,Harada24}. 
The interstitial vortices are still pinned 
even though they do not interact directly with the pinning sites,
since a periodic potential is created by the repulsive interaction of the
vortices trapped at the pinning sites. The interstitial vortices are typically 
far less strongly pinned than the vortices at the pinning sites, 
so that under an applied drive  
the interstitial vortices move while
the vortices at the pinning sites remain immobile.
Evidence for the motion of the interstitial vortices has been provided
by transport measurements \cite{Moschalkov23} 
as well as direct imaging in square pinning arrays, which 
show that the interstitial vortices move in 1D paths   
between the pinning sites \cite{Harada24}. 
In simulations, 1D 
interstitial vortex flow between pinning sites 
has also been observed \cite{DrivenShort28,Zhu29}.  
Recent experiments \cite{VanLook8} 
and simulations \cite{Shaprio29} have 
demonstrated the appearance of steps in the V-I
characteristics for moving
interstitial vortices when a DC and AC drive are superimposed.  
In the previous theoretical work for an applied transverse AC drive,
vortex filling fractions between $1.0$ and $2.0$ were examined, 
where each pinning site captures only one vortex \cite{Shaprio29}.     

In this work we examine the additional fractional steps
for the square pinning array with a transverse AC drive. We
find fractional steps both above and below the main step. 
We show that the width of  most of the fractional steps scales
as the square of the AC amplitude, as previously shown for the
depinning current and the main step \cite{Phase14}. 
Along the steps, the vortex orbits are stabilized in fixed trajectories,
while in the non-step regions, the vortex motion is chaotic in
appearance, producing a distinct velocity noise spectra. 
We also find scaling in the 
velocity versus applied DC 
drive at depinning and near the main step, with $\beta = 1/2$.
The steps are most prominent for commensurate vortex filling fractions
where the interstitial vortices form a symmetrical ground state. For
filling fractions where the vortices are disordered, the phase locking
phenomena is absent, and for filling fractions just off of a commensurate 
filling, the steps have a linear increase with drive due to the
fact that certain portions of the sample are not undergoing phase locking.
We find that even for small finite temperatures, the fractional
steps become completely washed out. The main step is visible up to the
melting point of the interstitial vortex lattice.  
With triangular pinning we
find a much more pronounced main step
and fractional steps, with the depinning threshold nearly the same
as that of the square case.
The widths of the main and most of the fractional
steps scale linearly with the AC amplitude for triangular pinning.    

\section{Simulation}

We numerically integrate the overdamped equations of motion, which 
for a single vortex $i$ is
\begin{equation}
{\bf f}_{i} = {\bf f}_{i}^{vv} + {\bf f}_{i}^{vp} + {\bf f}_{DC}  
+ {\bf f}_{AC} = 
{\bf v}_{i} \;
\end{equation}
The 
total force acting on vortex $i$ is 
${\bf f}_{i}$, ${\bf f}_{i}^{vv}$ is the 
total force from all the other vortices, ${\bf f}_{i}^{vp}$ is the 
force from the
substrate, ${\bf f}_{DC}$ is the force from the DC drive and  
${\bf f}_{AC}$ is the force from the AC drive. 
The simulation is in 2D with 
periodic boundary conditions in the $x$ and $y$ directions. 
For the square pinning case we place an 
array of $N_p = N \times N$ pinning sites in the sample, where
each pinning site is an attractive parabolic potential of radius $r_{p}$ and
a maximum force $f_{p}$. 
To create a triangular pinning array, we displace every other row of
pinning sites by half a pinning lattice constant.
The force from the pinning 
sites is $f_{i}^{vp} = \sum_{k}^{N_p}
(f_{p}/r_{p}) |{\bf r}_{i}-{\bf r}_{k}^{(p)}| \ 
\Theta(r_{p} -|{\bf r}_{i} - {\bf r}_{k}^{(p)}|)
{\hat {\bf r}}_{ik}^{(p)}$, 

\begin{figure}
\centerline{
\epsfxsize=3.5in
\epsfbox{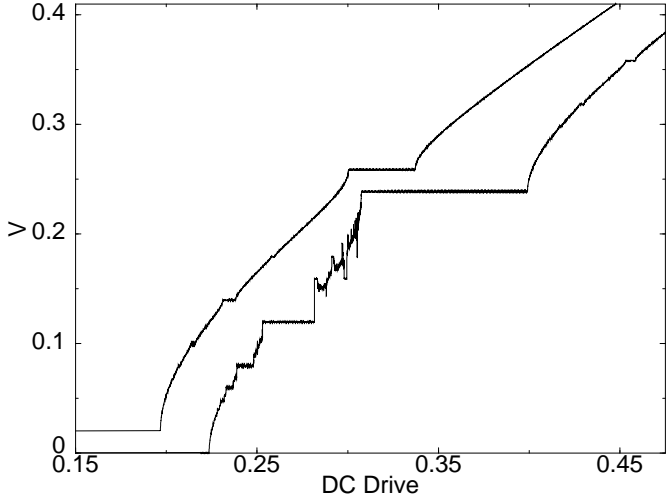}}
\caption{
Velocity $V$ vs applied DC drive $f_{DC}$ for AC amplitudes 
$A = 0.25$ (upper curve) and $0.35$ (lower curve) for a square pinning array. 
The upper curve has been offset upwards by 0.02 for clarity.
}
\label{fig1}
\end{figure}

\hspace{-13pt}
where $\Theta$ is the step function, 
${\bf r}_{k}^{(p)}$ is the location of pinning site $k$, 
and ${\hat {\bf r}}_{ik}^{(p)} = ({\bf r}_{i} - 
{\bf r}_{k}^{(p)})/| {\bf r}_{i} - {\bf r}_{k}^{(p)}|$. 
In our system each pin captures only one vortex.   
The total force from the vortex-vortex interaction is
${\bf f}_{i}^{vv} = -\sum_{j\neq i}^{N_{v}}\nabla_i U_{v}(r_{ij})$
where we 
use $U(r) = -ln(r)$, 
the Pearl vortex-vortex interaction potential,
appropriate for thin film superconductors.
The experiments with periodic pinning arrays are in this 2D limit. 
For computational
efficiency we use a summation technique \cite{Jensen30} to
evaluate the vortex-vortex interaction. 
Temperature can also be applied to the system by adding a noise
term $f_{i}^{T}$ to the equation of motion.
This noise term has the  
properties $<f_{i}^{T}(t)> = 0.0$ and
$<f_{i}^{T}(t)f_{j}^{T}(t^{'})> = 2\eta k_{B}T\delta_{ij}\delta (t - t^{'})$.
Here we set $\eta = k_{B} = 1$. 

We have considered 
pinning arrays from $4\times4$ to $10\times10$ pins and observe
the same results in each case.
The initial vortex configurations are obtained by annealing
from a high temperature where the vortices are in a liquid state 
and cooling to $T=0$. The DC driving is in the
$x$-direction, $ {\bf f}_{DC} = f_{DC}{\hat {\bf x}}$, 
and the AC force is in the
$y$-direction, ${\bf f}_{AC} =  A\sin(\omega t){\hat {\bf y}}$. 
The DC force is
increased from zero in increments of 0.0009. We average 
the vortex velocities $V = \sum {\bf v}_{i} \cdot {\bf \hat{x}}$ at each
increment for 175000 MD steps, and the resulting DC force
versus velocity curve
is proportional to the DC voltage-current curve.

\section{Fractional Steps for Square Pinning Arrays} 

In Fig.~1 we show the vortex velocity 
$V$ versus applied DC drive $f_{DC}$ for two different AC 
amplitudes, $A=0.25$ (upper curve) and $A=0.35$ (lower curve), for a fixed
frequency of $\omega=0.002$. 
Here there is a large step, centered near 
$f_{DC} = 0.32$ for $A=0.25$ and 
$f_{DC} = 0.35$ 

\begin{figure}
\centerline{
\epsfxsize=3.5in
\epsfbox{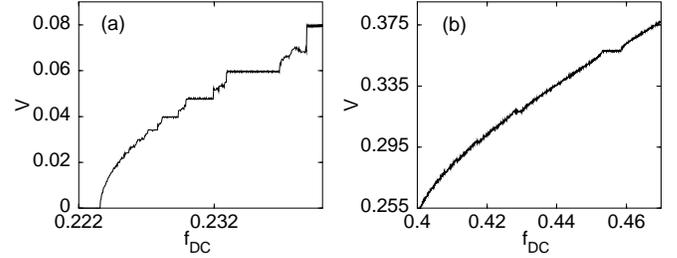}}
\caption{Magnified regions of vortex velocity $V$ versus
applied DC drive $f_{DC}$ for $A=0.35$.  (a) $0.222 < f_{DC} < 0.241$, 
showing phase locking steps below the main step. 
(b) $0.4 < f_{DC} < 0.45$,   
showing phase locking steps above the main step. 
}
\label{fig2}
\end{figure}

\hspace{-13pt}
for $A=0.35$, which
corresponds to a phase where the vortex orbit is 
commensurate with the pinning, forming a sinusoidal like orbit
as studied previously. We label this step the ``main step'' or 1:1 step.
In addition to this
step, a number of smaller or 
fractional steps are visible both above and below the main step.
For example, for $A=0.35$ a prominent step appears near $f_{DC}=0.27$,
and a small step is visible near $f_{DC}=0.46$.
The widths of {\it all} the steps, 
including the depinning threshold, are smaller
for the lower AC amplitude.  
The depinning threshold drops from $f_{DC}=0.196$ for $A=0.35$ to
$f_{DC}=0.190$ for $A=0.25$, while the width $\Delta$ of the main step
drops from $\Delta=0.09$ for $A=0.35$ to $\Delta=0.04$ for $A=0.25$.
In Fig.~2(a) we show a magnified region
of $ 0.222 < f_{DC} < 0.241$, 
near the depinning threshold, for the 
$A = 0.35$ curve in Fig.~1. 
Here the large number of fractional steps present are clearly visible,
with the step widths getting progressively smaller 
for lower $f_{DC}$. In Fig.~2(b) we show a magnified region of 
the $A = 0.35$ curve above the main step, from $0.4 < f_{DC} < 0.45$,
where a series of fractional steps appear. 
Again the step widths decrease as $f_{DC}$ is lowered toward the
main step.  

\subsection{Commensurate Vortex Orbits and Noise} 

On steps at small $f_{DC}$ values, a vortex spends more than one
AC cycle in a single plaquette, while on steps at larger $f_{DC}$, 
the vortex moves through more than one plaquette in a single AC
cycle.
In Fig.~3 we illustrate the vortex trajectories 
(lines) and the vortex positions (dots) for fixed $f_{DC}$ 
drives on and near the steps in the $V$ versus $f_{DC}$ curve
of Fig.~1 for $A = 0.35$. 
Fig.~3(a) shows the vortex trajectory for the step near $f_{DC} = 0.235$,
which is most clearly visible in Fig.~2(a). 
Here, a stable vortex orbit occurs in which the vortex spends two AC
cycles in one plaquette before moving over to the next plaquette to
repeat the sequence.
We observe similar behavior at the smaller steps for 
$f_{DC} < 0.235$, where in the stable orbits, the
vortex spends $n$ or $n + 1/2$ complete AC cycles in one plaquette. 
In Fig.~3(b), in the stable orbit for the step at $f_{DC} = 0.243$ 
in Fig.~1, the vortex spends $3/2$ of an AC cycle in each plaquette.  
In Fig.~3(c) we show the vortex

\begin{figure}
\centerline{
\epsfxsize=3.5in
\epsfbox{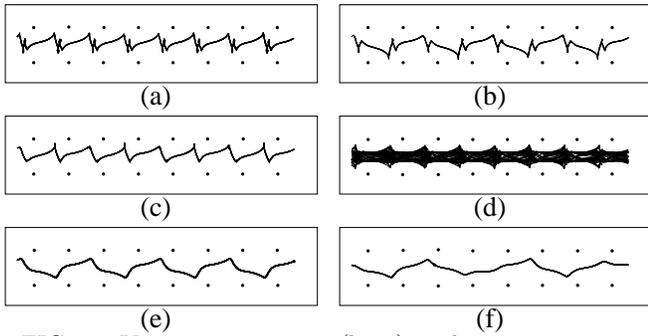}}
\caption{Vortex trajectories (lines) and vortex positions (black dots)
over equal time intervals
for constant $f_{DC}$ values chosen from the lower curve in Fig.~1 with
$A=0.35$.
The vortices in the square lattice are trapped in pinning sites and remain 
immobile; only the interstitial vortices are moving. 
The trajectory of only a single mobile vortex is presented for clarity. 
$f_{DC} = $ (a) $0.235$, (b) $ 0.243$, (c) $0.28$, (d) $0.305$, 
(e) $0.36$, and (f) $0.429$.}  
\label{fig3}
\end{figure}

\hspace{-13pt}
trajectory for $f_{DC} = 0.28$, which corresponds to the second largest
step in Fig.~1. Here the vortex goes through one AC cycle per plaquette. 
In the non-step region of Fig.~3(d), at $f_{DC} = 0.305$, 
there is no single stable vortex orbit.  Instead
the vortex trajectory wanders over time. There is an area close to the
occupied pinning sites where the vortex never flows due to the vortex-vortex 
repulsion. We find similar orbits for the other non-step regions. 
On the main step, illustrated in Fig.~3(e) at $f_{DC} = 0.36$,
the vortex moves through two plaquettes in a single AC cycle.
In Fig.~3(f) the vortex orbit for a step above the main step, 
$f_{DC} = 0.429$, has a similar pattern to that seen 
on the main step, Fig.~3(e), except that the vortex moves through
three plaquettes in each complete AC cycle.

The presence of stable orbits can also be inferred

\begin{figure}
\centerline{
\epsfxsize=3.5in
\epsfbox{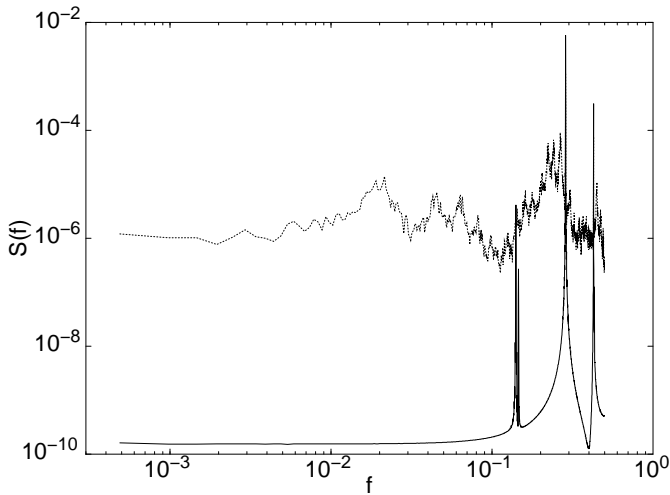}}
\caption{ 
Velocity noise $S(f)$ versus frequency $f$ for $A = 0.35$. Lower curve:
Noise signature along the main step corresponding to the vortex orbit at 
$f_{DC} =0.36$ shown in  Fig.~3(e). Upper curve: Noise 
signature in the non-step region for the vortex orbit
at $f_{DC}=0.305$ shown in Fig.~3(d).}   
\label{fig4}
\end{figure}

\begin{figure}
\centerline{
\epsfxsize=3.5in
\epsfbox{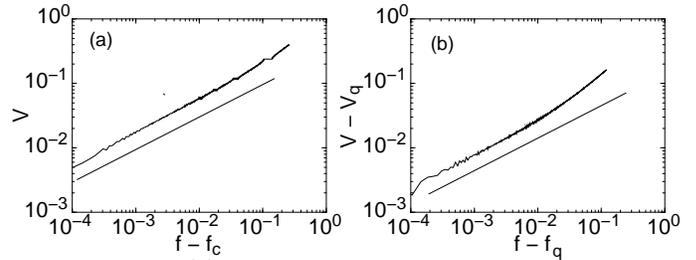}}
\caption{
(a) Log-log plot just above depinning of $V = (f -f_{c})$
for $A=0.35$, where  $f_{c}=0.196$ is the depinning
threshold and $f$ is the DC driving force. The solid line shows
a slope of $\beta = 1/2$. (b) Log-log plot above the main step of 
$V -V_{q} = (f -f_{q})$ where $V_{q}=0.24$ is the velocity along the main
step and $f_{q}=0.395$ is the DC driving force where the step ends. The
solid line shows a slope of $\beta = 1/2$.} 
\label{fig5}
\end{figure}

\hspace{-13pt}
through measurement of the velocity noise, which 
corresponds experimentally 
to the voltage noise. We acquire a time series of the velocity
in the $x$-direction for a fixed $f_{DC}$ value.  The 
corresponding noise spectrum is given by
$S(f) = |\int V(t)e^{-i2\pi f t} dt|^2$.
In Fig.~4, we 
examine the noise spectra for $A = 0.35$ for a step region, $f_{DC} = 0.36$, 
and a non-step region, $f_{DC} = 0.305$, corresponding to the orbits
in Fig.~3(e) and Fig.~3(d). 
In the step region (lower curve), a strong periodic signal 
is present at the frequency induced by the
AC driving, as indicated by the narrow large spike and the higher harmonics. 
For the non-step region (upper curve) there is is a broadened peak near the
AC driving frequency. The noise power for the lower frequencies is 
over 4 orders of magnitude higher than for the step region.
This lower frequency noise power comes from the vortex orbits drifting
over a time scale that is longer than the AC driving time scale.
The noise spectra does not have a broad band or $1/f^{\alpha}$ characteristic.
We find a similar noise spectra for the other non step regions. 

\subsection{Scaling Velocity vs Drive and Fractional Steps}

The velocity curves in Fig.~1 show considerable curvature near the 
depinning threshold and just above each step. 
Fisher has suggested that depinning
can be considered a dynamical critical phenomena \cite{Fisher16}
where 
\begin{equation}
V = (f - f_{c})^{\beta},
\end{equation} 
where $f$ is the DC driving force, $f_{c}$ is the driving force at
which depinning occurs, and $V$ is the velocity of the driven media.
In addition, in numerical work on CDW systems a similar scaling
is found near the mode locked steps \cite{Middleton17,Higgins18} in the form
\begin{equation}
|V - V_{q}| = |f -f_{q}|^{\beta} ,
\end{equation}
where $V_{q}$ is the velocity along the $q$th step and $f_{q}$ is the 
driving force at which the step ends or begins. For a single particle
moving in a 1D periodic substrate, $\beta = 1/2$. 
In charge density wave systems
with random pinning, 

\begin{figure}
\centerline{
\epsfxsize=3.5in
\epsfbox{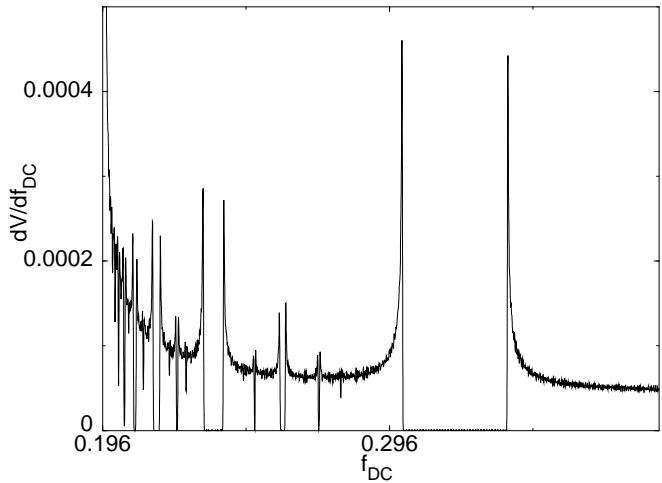}}
\caption{
The $dV/df_{DC}$ curve for $A = 0.35$ starting at the DC depinning
threshold of $f_{DC}=0.196$.} 
\label{fig6}
\end{figure}

\hspace{-13pt}
simulations of the depinning and step transitions 
give $\beta = 0.67$ in 2D and $0.83$ for 3D, while experimental
results are consistent with the results for the 3D regime 
\cite{Higgins18}. 
In our system, one might expect to find scaling with $\beta=1/2$ 
as in the 1D periodic substrate picture since
the vortices are commensurate with the periodic pinning.
On the other hand, the orbits in Fig.~3 indicate that
the vortices have a 2D velocity component and do not flow
in strictly 1D channels, so it is not clear if or how the
velocities will scale. 

In Fig.~5(a) we plot on a log scale
$V = f - f_{c}$ just above depinning for $A = 0.25$,
with the depinning threshold of $f_{c}=0.196$, 
and in Fig.~5(b) we plot $|V -V_{q}| = f - f_{q}$ 
above the main step for $A = 0.25$, with the velocity
$V_{q}=0.24$ along the main step, and where the end of the
step

\begin{figure}
\centerline{
\epsfxsize=3.5in
\epsfbox{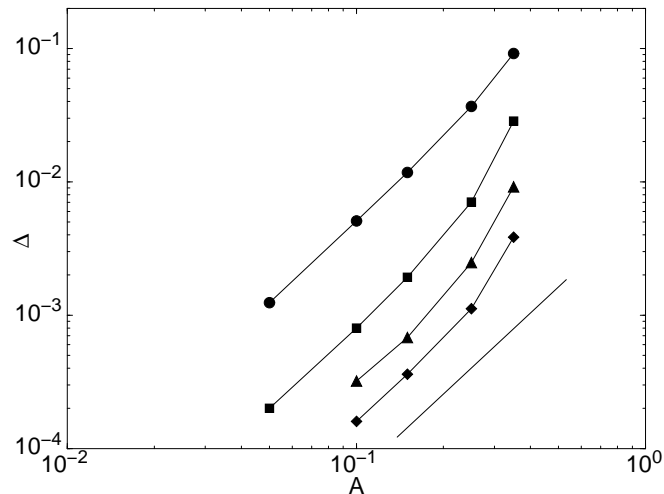}} 
\caption{The scaling of the step width $\Delta$ as a function of 
AC amplitude $A$ for the
fractional steps corresponding to: (diamonds) $f_{DC}=0.235$ [Fig.~3(a)],
(triangles) $f_{DC}=0.243$ [Fig.~3(b)], and (squares) $f_{DC}=0.28$ 
[Fig.~3(c)].  The scaling of the main 
step at $f_{DC}=0.36$ [Fig.~3(e)] is also shown (circles). 
The solid line is a quadratic fit.} 
\label{fig7}
\end{figure}

\begin{figure}
\centerline{
\epsfxsize=3.5in
\epsfbox{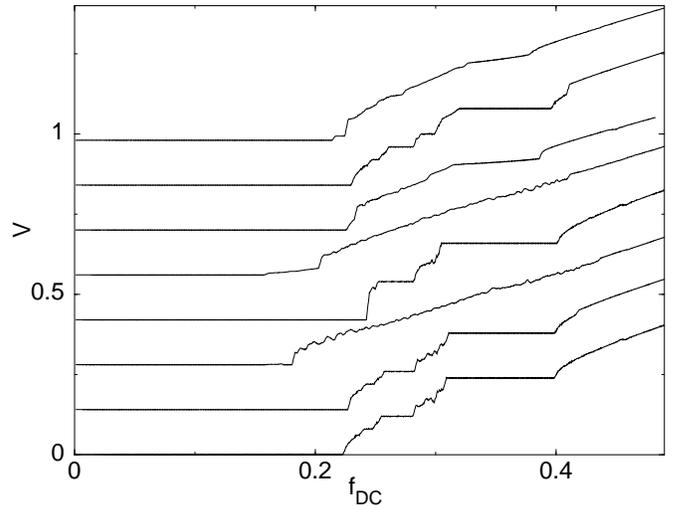}}
\caption{The $V$ vs $f_{DC}$ curves for different vortex filling 
fractions. The curves have been shifted up from $V = 0.0$ for 
presentation. 
From lowest to highest: $B/B_{\phi} =$ $17/16$, $5/4$, $0.34$, $3/2$, $0.67$,
$1.97$, $2.0$, and $2.04$. Here $A = 0.35$.} 
\label{fig8}
\end{figure}

\hspace{-13pt}
is at $f_{q}=0.395$.
Here in both
cases for low $f$ the curves fit well to $\beta = 1/2$, with a deviation 
to linear behavior for higher $f$. Our velocity resolution was too low 
to obtain adequate scaling for steps other than the main step. 
We find similar scaling for the 
depinning and the main step for all filling fractions at which
the interstitial vortices form a symmetrical configuration. For 
non-symmetrical vortex configurations, the scaling breaks down 
due to the fact that the depinning process 
for the non-symmetrical configurations is plastic, rather
than elastic as in the case of the symmetrical configurations. 
Scaling may still occur for the incommensurate  
systems as found in simulations with random pinning 
where plastic flow occurs, where scaling in the velocity versus 
drive was observed \cite{Dominguez31}
with $\beta > 1.5$.
Very large systems
beyond the scope of this work 
would be needed in order to determine if such 
scaling occurs for the incommensurate cases. 
In addition, some of the scaling breaks down for high AC amplitudes, as
seen for $A = 0.35$ in Fig.~1 where there are sharp jumps 
into and out of certain steps. 
For example, in 
Fig.~6 we plot the $dV/df_{DC}$ curve for $A = 0.35$, showing 
peaks going into and out of the phase locked regions. 
For $\beta = 1/2$ scaling, the overall shape of $dV/df_{DC}$ should 
scale with a power $-1/2$, which 
is consistent with the curve
in Fig.~6.  

We next show how the widths of different steps scale with AC amplitude.
In earlier work it was shown that the increase in the depinning force and
the width of the main step increase as the square of AC amplitude.
In Fig.~7
we plot the step widths $\Delta$ vs AC amplitude $A$
corresponding to the steps at $f_{DC}=0.235$ [Fig.~3(a)],
$f_{DC}=0.243$ [Fig.~3(b)], and 
$f_{DC}=0.28$ [Fig.~3(c)].
We also show the scaling for the main step, 
$f_{DC}=0.36$ [Fig.~3(e)], for a comparison.
The widths all fit well to $A^2$ with some deviations at higher AC amplitudes,
where the widths increase faster than $A^2$.

\begin{figure}
\centerline{
\epsfxsize = 3.5in
\epsfbox{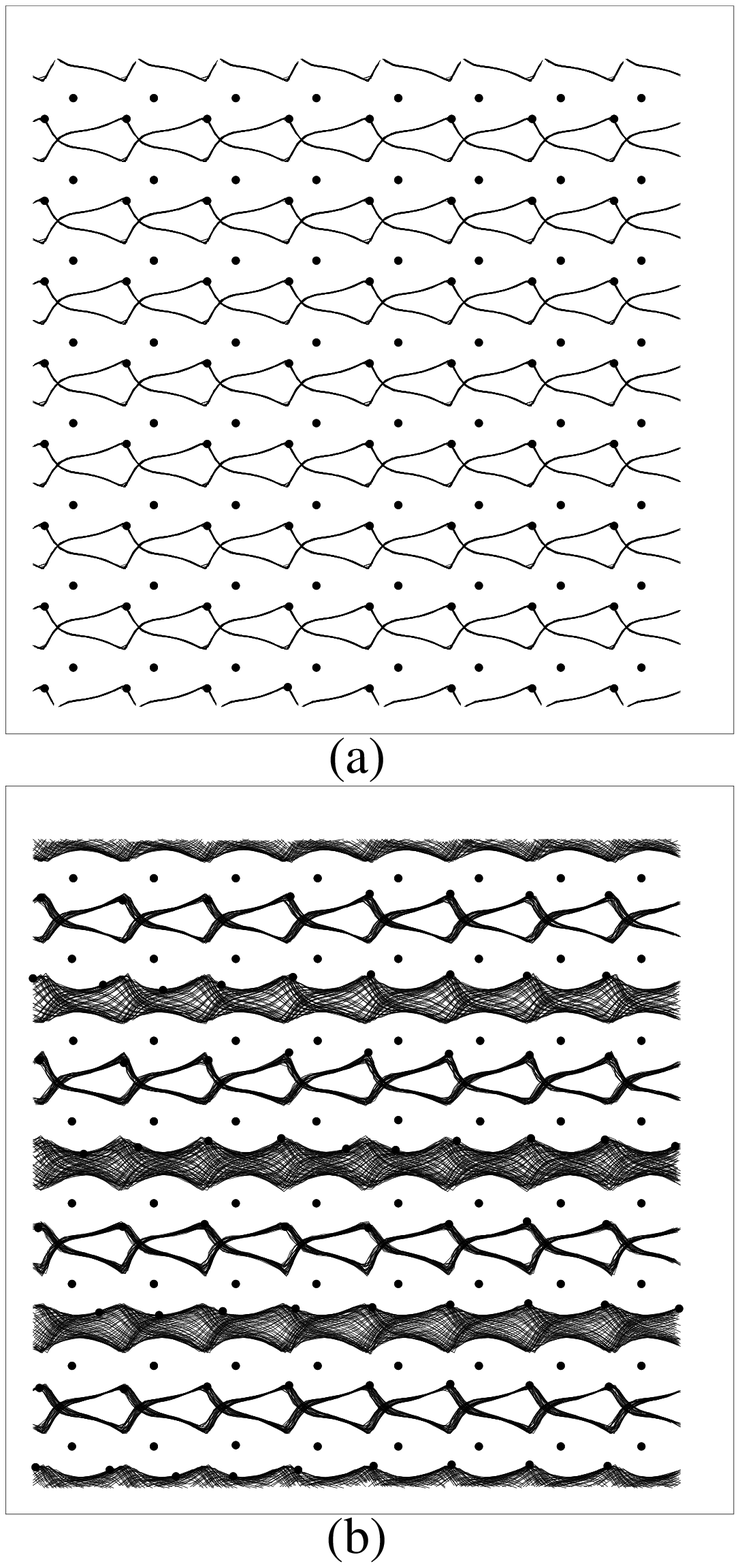}} 
\caption{
Vortex trajectories (lines) and positions (black dots) 
along the main step for $f_{DC}=0.36$ and $A=0.35$ for (a) 
$B/B_{\phi} = 2.0$, showing complete phase locking; 
(b) $B/B_{\phi} = 2.04$, showing 
that only a certain portion of the
moving vortices are undergoing phase locking.} 
\label{fig9}
\end{figure}

\subsection{Effect of Filling Fraction}

In Fig.~8 we show a series of velocity vs $f_{DC}$ curves for increasing
filling fractions from (bottom to top)
$B/B_{\phi} =$ $17/16$, $5/4$, $0.34$, $3/2$, $0.67$, $1.97$, $2.0$,
and $2.04$. 
The interstitial vortex lattice forms an ordered symmetrical ground state for
$B/B_{\phi} =$ $17/16$, $5/4$, $3/2$, and $2.0$. At these fillings the 
phase locking steps are the most pronounced, and the velocity
curves are almost
identical for each one. 
These are also the same filling fractions where peaks
in the critical current are observed in experiments and simulations.
Since the interstitial vortices form a symmetrical ground state,
the effective interstitial vortex interactions cancel, 
and the system can be thought of as
a single particle moving through a periodic substrate. In the absence
of an AC drive, the vortices collectively move in 1D paths under a DC drive.  
For the 
filling fractions $B/B_{\phi}=0.34$ and $0.67$, 
the ground state is disordered and the phase locking is absent. 
In the absence of an AC drive the vortices move plastically in winding paths.  
Under an applied AC drive, this winding motion destroys any phase locking.

For filling fractions $B/B_{\phi}=1.97$ and $2.04$,
just below and just above the commensurate filling
at $B/B_{\phi}=2.0$, the phase locking still occurs; however, the
steps have an additional linear velocity increase superimposed on them. 
In Fig.~9(a) we illustrate
vortex trajectories along the main step for $B/B_{\phi} = 2.0$ and 
in Fig.~9(b) for $B/B_{\phi} = 2.04$. For $B/B_{\phi} = 2.0$ the vortices
move in well defined stable sinusoidal trajectories, while
for $B/B_{\phi} = 2.04$ only certain rows of vortices have this ordered
motion while the vortices in the other rows have unstable ordering. Only the
rows with stable orbits are phase locked and remain at the same 
velocity along the step for increasing $f_{DC}$. 
The vortices moving through the unstable channels are not 
phase locked and therefore their velocity increases
with $f_{DC}$, accounting for the linear velocity increase in the
$V$ - $f_{DC}$ curves along the step. 
The stable moving vortex rows have a commensurate
number of vortices. 
In the system presented here, for an $8\times8$ pinning lattice
there are eight interstitial vortices in each commensurate row. 
There are more than eight vortices in the rows with unstable vortex 
trajectories, so the rows are incommensurate. For certain filling fractions
such as $B/B_{\phi} = 0.67$, all the rows are incommensurate and there
is no phase locking. 
For $B/B_{\phi}=1.97$, where the unstable or
incommensurate rows have {\it less} than eight vortices,
we observe vortex trajectories similar to those in Fig.~9(b).
We note that similar
behavior occurs just above and below the filling fractions of $5/4$ and
$3/2$.  These results are consistent with experimental results in
which Shapiro steps were most 
clearly defined at $B/B_{\phi} = 2.0$ \cite{VanLook8}.   

\subsection{Effect of Temperature}

We consider the effects of temperature by adding a noise term
$f_{i}^{T}$ to the
equation of motion for the vortices. 
We normalize our temperature by 
the melting temperature $T_{m}$, where $T_{m}$ is the temperature at which
the vortices began to diffuse in the absence of an external drive. In our 
system this onset is sharp and well defined, due 

\begin{figure}
\centerline{
\epsfxsize = 3.5in
\epsfbox{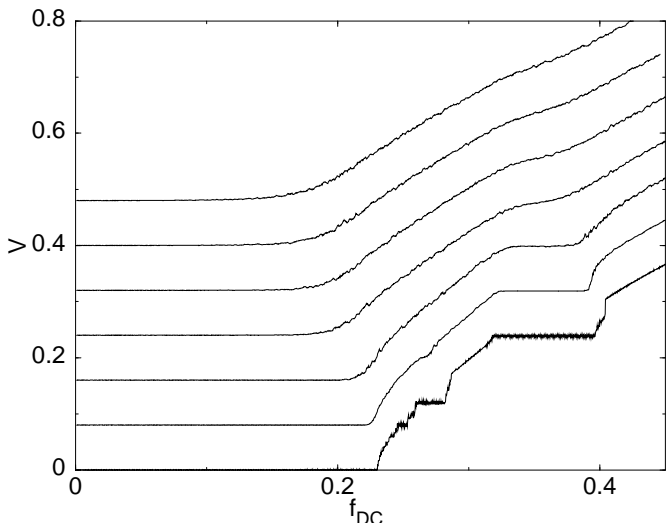}}
\caption{
The $V$ vs $f_{DC}$ curves for different temperatures for $B/B_{\phi} = 2.0$
and $A = 0.35$. The curves have been shifted up from zero for presentation.
From lowest to highest, $T/T_{m} =$ $0.0$, $0.07$, $0.28$, $0.64$, $0.75$, 
$0.87$, and $0.96$, where $T_{m}$ is the melting temperature.} 
\label{fig10}
\end{figure}

\hspace{-13pt}
to the fact that the
vortices are sitting in a periodic substrate. 
In the absence of a substrate the 
onset of diffusion is more gradual and a well defined melting temperature
does not appear.  Experimentally, the melting of the interstitial vortices
occurs at or just below $T_{c}$ 
\cite{Metlushko21,Moschalkov23} so that $T_{m} \approx T_{c}$.   
In Fig.~10 we show a series of $V$ vs $f_{DC}$ curves
for $A = 0.35$ for
$T/T_{m} =$ $0.0$, $0.07$, $0.28$, $0.64$, $0.75$, $0.87$, and $0.96$, 
from bottom to top. 
The fractional steps wash out very quickly with 
temperature at about $T/T_{m} = 0.1$.  The main step is visible all the way up
to $T_{m}$ but is more cusp like for $T/T_{m} > 0.64$. This washing out of the
step with temperature is also consistent with the Shapiro step experiments
near $T_{c}$ which found only a cusp feature
\cite{VanLook8}. 
Experimentally it would still
be possible to measure the increase 
of the step width with AC amplitude by taking
the derivative of the $I-V$ curves, which shows a dip at each 
side of the step.
Since most transport experiments are performed close to $T_{c}$, 
it will be very difficult to observe
the fractional steps. The widths of the steps can be increased 
by increasing $A$,
allowing fractional steps to be visible for $T > 0.1T_{m}$.
In practice, however, 
there will be a limit to how large $A$ can be made before the vortices
at the pinning sites begin to depin.

\section{Transverse Phase Locking For Triangular Pinning}

Next we study the transverse phase locking for the case where the pinning 
is triangular rather than square. We use the same system size and pinning 
density as for the square pinning array, but
every other pinning row is shifted in the $x$-direction by half a lattice
constant. Here the motion of an interstitial vortex for zero AC amplitude 

\begin{figure}
\centerline{
\epsfxsize = 3.5in
\epsfbox{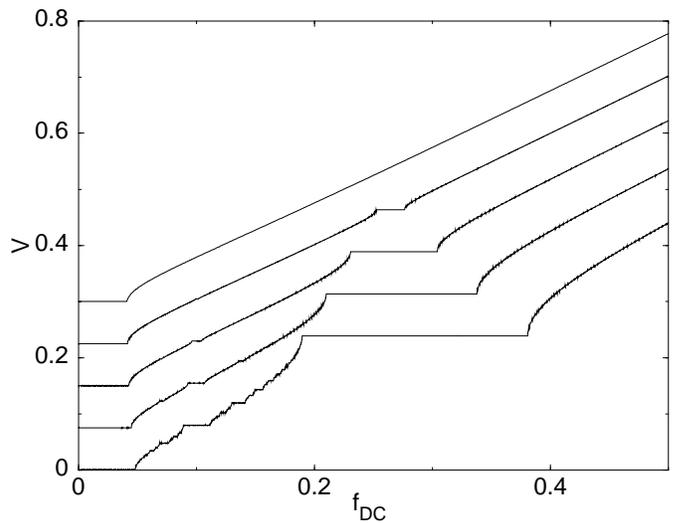}}
\caption{ The velocity $V$ vs DC driving force 
$f_{DC}$ curves for a triangular pinning array
at different AC amplitudes $A$. 
The system size and pinning density are the same as
for the system in Fig.~1. The curves have been shifted up for 
presentation. From lowest to highest, $A = 0.35$, $0.25$, $0.15$, 
$0.05$, and $0.0$.}   
\label{fig11}
\end{figure}

\hspace{-13pt}
has a periodic transverse velocity component, unlike the square case where 
for zero AC amplitude the motion is strictly 1D. 
In Fig.~11 we show
$V$ vs $f_{DC}$ 
for different values of $A$. We find phase locking similar to that
seen in the square pinning case.  The step widths all increase monotonically 
with AC drive, rather than oscillating as expected for Shapiro steps. 
There are some noticeable differences 
from the square case, in that the depinning force is only 
weakly affected by increases in $A$. In addition the main step is 
wider than that observed for square pinning at the same AC amplitudes.
We also find that the step widths increase linearly
with AC amplitude, rather the quadratically as
in the square pinning array.      

In Fig.~12(a) we show the vortex motion above depinning for $A =0.0$,
indicating that
the vortex moves with a periodic transverse component even in the absence
of an AC drive. 
In Fig.~12(b-f) we show the vortex orbits along various steps for 
$A = 0.25$. 
In Fig.~12(b), at $f_{DC} = 0.07$, a stable
periodic orbit forms
where the vortex goes through 
an interesting loop that alternates from up to down.  
This loop feature appears for stable orbits where $f_{DC} < 0.1$. 
For the step near $f_{DC} = 0.11$ [Fig.~12(c)], 
corresponding to the second largest step,
a stable orbit forms similar to that in Fig.~12(b) but the loop
feature is absent.
On the small step at $f_{DC} = 0.1375$ [Fig.~12(d)],
a stable orbit forms which is not symmetrical in the $y$-direction. 
A vortex trajectory for a non-step region at $f_{DC}=0.2$
[Fig.~12(e)] has similar characteristics as the trajectories
for the non-step regions for square pinning. In Fig.~12(f) we show the
vortex orbit along the main step, $f_{DC} = 0.3$, where a well 
defined sinusoidal orbit can be seen as the transverse frequency 
generated by the DC motion matches one-to-one with the frequency of 

\begin{figure}
\centerline{
\epsfxsize = 3.5in 
\epsfbox{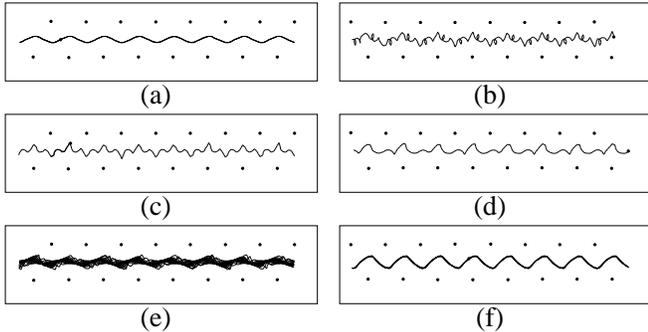}}
\caption{Vortex trajectories 
for triangular pinning on various steps for
(a) $A = 0.0$ and  (b-f) $A = 0.25$.
(a) For $A = 0.0$ above depinning, $f_{DC}=0.3$,
there is a periodic transverse motion of
the vortex even for zero AC drive. 
Remaining panels: $A=0.25$ and 
(b) $f_{DC} = 0.07$, (c) $f_{DC}=0.11$, (d) $f_{DC}=0.1375$, 
(e) $f_{DC}=0.2$, and (f) $f_{DC}=0.3$.}   
\label{fig12}
\end{figure} 

\hspace{-13pt}
the transverse AC drive. For higher DC drives, there 
are additional stable
orbits along the steps as in the case for the square pinning.  
Similar noise signals for step and non-step regions as seen in Fig.~4
are also observed for the triangular pinning case.

In Fig.~13 we show the scaling of the width $\Delta$ of the main step
for triangular pinning with AC drive $A$,
along with the width $\Delta$ of the main step for square pinning
for comparison.
Here the magnitude of $\Delta$ for the triangular pinning is greater than
that for the square pinning for all AC values considered here. The width 
of the triangular steps also scales linearly with AC amplitude rather
than quadratically as in the square pinning, so that for smaller AC amplitudes
the difference in the magnitudes of $\Delta$ becomes greater.  
This result suggests that the detection of transverse phase locking would
be easiest for a system with {\it triangular} 
rather than square pinning arrays.

\begin{figure}
\centerline{
\epsfxsize = 3.5in
\epsfbox{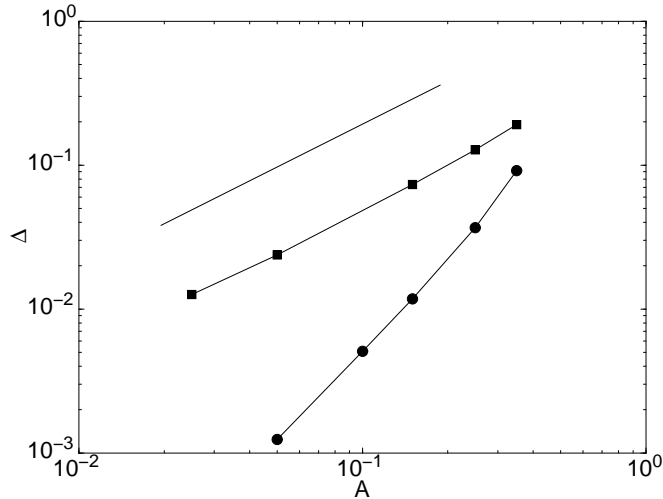}}
\caption{ Scaling of the step width $\Delta$
versus AC amplitude $A$ of the main step for
triangular pinning (squares), along with 
the scaling for the main step for square
pinning (circles) for comparison. The solid line is a linear fit.} 
\label{fig13}
\end{figure}

\begin{figure}
\centerline{
\epsfxsize = 3.5in
\epsfbox{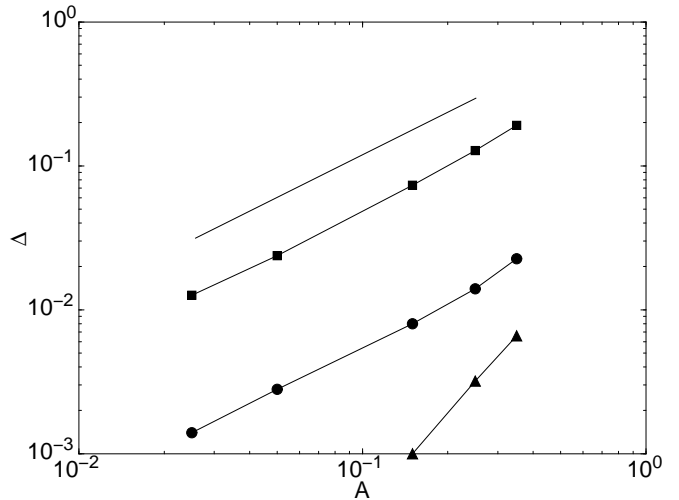}}
\caption{Scaling of the step width $\Delta$
vs $A$ for triangular pinning 
for (circles) $f_{DC}=0.11$ [Fig.~12(c)], along
with (triangles) the depinning force.
We also show the scaling of the main step (squares) for comparison.
}
\label{fig14} 
\end{figure}

In Fig.~14 we show the scaling of the step width $\Delta$ with
AC amplitude $A$ for the fractional step at $f_{DC}=0.11$ (the orbit shown in
Fig.~12(c)), along with the width of the main step and the
change in value of the depinning force.
Here the fractional step scales linearly with 
AC amplitude, just as the main step does. 
The difference in the depinning force at different values of $A$, which 
is much smaller than the widths of the main step and the fractional step, 
does not scale linearly with $A$ but increases quadratically or even faster.  

We can compare our results for 
transverse phase locking in the triangular pinning 
to the results for random pinning. 
For random pinning \cite{Kolton215}
the vortices in the highly driven phase regain partial triangular
ordering and a washboard signal appears. Unlike the square pinning case, 
in random pinning in the absence of a transverse AC drive the 
vortices do not move in strictly 1D channels
but have a transverse velocity 
component centered around zero which can also have a washboard signal. 
For the random pinning
it was found that the width of the transverse phase locking step
oscillates as
a function of AC amplitude, 
which is different from the results for the triangular 
or square pinning.  For low AC amplitudes, however, the increase in the
step width is {\it linear} with AC amplitude in the random
pinning case.  Using a simple model it 
was shown that any form of disorder that induces a transverse temporal order
in the absence of an AC drive 
gives rise to a linear dependence of the step width 
for small AC amplitudes 
\cite{Kolton215}. 
This is similar to the case of 
the triangular pinning, where there is transverse temporal order
present in the zero AC drive limit. In addition, for the case of
random pinning, the vortex trajectories
along the step show stable sinusoidal orbits
in a similar manner to the stable orbits found in the periodic 
pinning cases.    

\section{Conclusion} 
We have investigated the transverse phase locking in square and 
triangular pinning arrays. For square arrays we find a series of 
fractional steps which correspond to stable vortex orbits.
Along the non-step regions there are no stable periodic vortex orbits.
 All the
fractional steps increase monotonically in width with AC amplitude. 
Along the steps a narrow band noise signal is present, while in the
non-step regions the narrow band signal is broadened and 
large low frequency noise power appears.
We have shown 
explicitly that the width of several of the fractional steps increases as the  
the square of the AC amplitude. The
velocity vs drive near the depinning threshold and the 
main step scale with $\beta = 1/2$.    

We find that the phase locking is most pronounced for interstitial vortex
filling fractions at which the interstitial vortices form a symmetrical ground
state, such as at $B/B_{\phi} = 1.25, 1.5$ and $2.0$. For filling 
fractions near these commensurate fillings, a partial phase locking occurs
where certain regions of the sample have stable phase locked orbits while
other regions are unstable and the vortex velocity in these regions increases
linearly. For filling fractions where the ground states are disordered the
phase locking is absent. With a finite temperature, the fractional steps
appear only at low temperatures and are washed out at higher temperatures,
while the presence of the main step can
be detected up to the melting temperature $T_{m}$. 

For triangular pinning arrays, where moving interstitial vortices have
a periodic transverse motion even in the absence of a transverse AC drive,
we again observe a transverse phase locking with all the step
widths increasing
monotonically with AC amplitude rather than oscillating as in the case 
of Shapiro steps. 
The depinning force is only weakly increased by increasing the AC
amplitude.
The main step for the triangular case is larger
than that observed in the square pinning case, and its width 
increases linearly with AC amplitude, 
unlike the quadratic increase observed for the square case.
The linear increase in the step widths with AC amplitude is similar
to the results for transverse phase 
locking with random disorder with small AC drives, where
there is also an ordered transverse motion in the absence of an AC drive. 
We show that the width of the fractional steps also increase linearly 
with AC amplitude,
but the difference in the depinning force increases quadratically.   

Our predictions should be testable for superconductors with periodic
pinning arrays where only one flux line is captured per pinning site. 
The steps can best be observed in samples where Shapiro steps have
already been observed with longitudinal DC and AC drives at filling fractions
where a symmetrical vortex configuration occurs, such as 
$B/B_{\phi}=17/16$, $5/4$, $3/2$, or
$2$. For experiments performed near $T_{c}$, it is unlikely that the fractional
steps can be observed; however, the main steps should be visible.  
A particular advantage of transverse  phase locking steps
over longitudinal (or Shapiro) steps 
is that the step width can be made arbitrarily 
large by increasing the AC amplitude or lowering the AC frequency. 
Our results should also be relevant for vortex motion in Josephson-junction
arrays at commensurate fillings. 

{\bf Acknowledgments:}
We thank D. Dom{\' \i}nguez, N. Gr{\o}nbech-Jensen, 
A. Kolton, V. Moshchalkov, D. Stroud, and
L. Van Look for
useful discussions.
This work was supported by the US Department of Energy under
contract W-7405-ENG-36.

\end{document}